\RequirePackage{fix-cm}
\documentclass[smallcondensed]{svjour3}     

\smartqed  

\usepackage{graphicx}
\usepackage{url}
\usepackage{booktabs}
\usepackage{multirow}
\usepackage{subfig}
\usepackage{amsmath}

\usepackage{caption}

\usepackage{breakurl}
\usepackage{hyperref}
\usepackage{color,soul}
\usepackage{todonotes}
\newcommand {\mudvisualizer }{MUD-Visualizer}
\newcommand {\mudfile }{MUD-File}
\newcommand {\mudfiles }{MUD-Files}

\begin{document}

\title{Making Access Control Easy in IoT 
}


\author{Vafa Andalibi         \and
        Jayati Dev  \and
        DongInn Kim \and
        Eliot Lear \and
        L. Jean Camp
}


\institute{V. Andalibi, J. Dev, D. Kim, L. J. Camp \at
              Indiana University, Bloomington, IN, USA \\
              \email{\{vafandal, jdev, dikim, ljcamp\}@indiana.edu}           
           \and
           E. Lear \at
            Cisco Systems, Zurich, Switzerland \\
              \email{lear@cisco.com}   
}

\date{Received: date / Accepted: date} 



\maketitle
  
\begin{abstract}
Secure installation of Internet of Things (IoT) devices requires configuring access control correctly for each device. In order to enable correct configuration the Manufacturer Usage Description (MUD) has been developed by Internet Engineering Task Force (IETF) to automate the protection of IoT devices by micro-segmentation using dynamic access control lists. The protocol defines a conceptually straightforward method to implement access control upon installation by providing a list of every authorized access for each device. This access control list may contain a few rules or hundreds of rules for each device. As a result, validating these rules is a challenge. In order to make the MUD standard more usable for developers, system integrators, and network operators, we report on an interactive system called \mudvisualizer{} that visualizes the files containing these access control rules. We show that, unlike manual analysis, the level of the knowledge and experience does not affect the accuracy of the analysis when \mudvisualizer{} is used, indicating that the tool is effective for all participants in our study across knowledge and experience levels. 


\end{abstract}

\section{Introduction}
\label{sec:intro}

The forecast for the number of connected IoT devices in 2025 is now raised to 30.9 billion \cite{iotLueth2020}, yet their (in)security is still a major concern 
 \cite{o2016insecurity,mangino2020internet,thilakarathne2020security}.
There is a need for secure onboarding meaning that the device is secured as soon as it is connected to the network. One major component of secure onboarding both for cyber-physical systems and IoT is firewall configuration. Without access control, IoT devices are susceptible to participate in DDoS attacks \cite{kolias2017ddos}, are vulnerable to ransomware \cite{yaqoob2017rise}, and enable information exfiltration from within networks \cite{d2016data}. 
It is the nature of botnets that the subverted devices need to be controlled by the attackers' command and control (C2) infrastructure 
\cite{bailey2009survey}.
Secure onboarding that implements allow-list access control limits exposure of devices to attacks and prevents any subverted device from connecting to the attackers' C2 points. Unlike traditional botnets, the control servers in IoT are highly dynamic so the typical response of identifying then block-listing is infeasible \cite{tanabe2020disposable}. 

To this end, the Internet Engineering Task Force (IETF) has developed the Manufacturer Usage Description (MUD); a standard that provides an isolation-based defense for IoT devices using dynamic access control \cite{rfc8520}. The urgency and scale of the need for such a solution are shown by the fact that MUD is also a part of the National Institute of Standard and Technology (NIST) security for IoT initiatives \cite{dodson2019securing}. In addition, the Department of Commerce has a working group to integrate the Software Bill of Materials (SBoM) initiative with MUD\footnote{https://www.ntia.doc.gov/files/ntia/publications/ntia\_practices\_model\_and\_summary\_19-02-20\_0.pdf}
and the IETF has a proposed standard integrating SBoM with MUD\footnote{https://tools.ietf.org/html/draft-lear-opsawg-mud-sbom-00}. 
MUD can also be used for mitigating DDoS attacks in the Fog \cite{andalibi2019throwing}. 

MUD relies on manufacturers for an Access Control List (ACL) in the form of a \mudfile.  A \mudfile{} defines the allowed and expected behaviors of the associated device. 
The clear implication is that developers must be able to write clear and correct \mudfile{}s and network operators must be able to read and validate the \mudfile{}s to ensure that unnecessary communications, either locally or over the Internet, are not allowed. These are difficult problems, and like many security tasks, are not well aligned with human cognitive abilities \cite{oliveira2014s}.

In this work, we report on the usability analysis of the \mudvisualizer{} \cite{andalibi2021mudviz}; a tool that is intended to support developers and network operators in evaluating overlaps, duplication, and possible conflicts in \mudfile{}s. We report on the design and results of our human subjects research that we conducted to investigate the following research questions: 

\begin{description}
    \item \textbf{RQ1}: How does \textbf{security knowledge} affect the accuracy of the analysis of the \mudfiles{}? 
    \item \textbf{RQ2}: How does \textbf{security experience} affect the accuracy of the analysis of the \mudfiles{}? 
    \item \textbf{RQ3}: To what extent does \textbf{level of knowledge and experience} affect the accuracy of the analysis of the \mudfiles{}? %
\end{description}
\section{The MUD Standard}
\label{sec:mud}
In this section, we briefly review the MUD standard for those readers who are unfamiliar with MUD. MUD is comprised of six main components: \textbf{\mudfile{}} which is a YANG-based JSON file (RFC 7951) created and digitally signed by the manufacturer. It embeds the behavioral profile of the IoT device in an access control list. \mudfile{}s should be hosted on manufacturer's \textbf{MUD file server}. The location of these files on the Internet is the \textbf{MUD-URI} which is stored on the IoT device. Upon connection of the device to a MUD-compliant network, the device sends the embedded MUD-URI to the Authentication, Authorization, and Accounting, i.e., \textbf{AAA server}. The \textbf{MUD-Manager} is the core of MUD architecture. After receiving the MUD-URI, it will retrieve the \mudfile{} from the manufacturer's MUD file server and communicates the \mudfile{} rules to the AAA server \cite{rfc8520}. The \textbf{Network Access Device (NAD)} (i.e., the router) is equipped with an internal firewall that is configured by the AAA server. 
MUD provides seven abstractions that can be used to define the behavior of and constraints on an IoT device in a \mudfile{}. The \textbf{domain-name} abstraction is used to enforce restrictions on cloud access. The \textbf{local-networks} abstraction defines the communication of a device with other devices on the network. With the \textbf{manufacturer} abstraction, the authority component (i.e., domain name) of a device is matched against the MUD-URI of another node which restricts devices' access to specific manufacturers. Similarly, the \textbf{same-manufacturer} abstraction defines when devices built by one manufacturer can communicate with each other but not with devices built by other manufacturers. Both of the \textbf{controller} and \textbf{my-controller} abstraction are used when devices use a controller to communicate.  Lastly, the \textbf{model} abstraction constrains a device to communicate only with other instances of the same device (e.g., only lightbulbs interact) \cite{rfc8520}.

To address the human factors challenges in the analysis of the \mudfile{}s, Andalibi et al. \cite{andalibi2021mudviz} proposed and implemented \mudvisualizer{} with the goal of 1) protocol checking to avoid formatting errors in the \mudfile{} to prevent coding errors 2) identifying internal inconsistencies and inefficiencies to prevent logic errors  
3) enabling both manufacturers and sysadmins to review and validate the \mudfiles{} by processing the abstractions' access control rules and visualizing them. This processing is performed by encoding the merged Access Control Entries (ACEs) into a tree (i.e., ACE Tree) followed by pruning that tree to remove the duplicate ACEs that are generated by merging the MUD abstractions in two or more \mudfiles{} \cite{andalibi2021mudviz}. \mudvisualizer{} can be deployed either as a stand-alone app or as a web app. It is scalable, open-source, and publicly available online on GitHub \cite{andalibi2021mudviz}.

\section{Related Work}
\label{sec:related}


Currently there are five implementations of MUD: Cisco MUD\footnote{https://github.com/CiscoDevNet/MUD-Manager}, NIST MUD\footnote{https://tsapps.nist.gov/publication/get\_pdf.cfm?pub\_id=927289}, osMUD\footnote{https://github.com/osmud/osmud}, Masterpeace MUD (closed-source), and CableLabs Micronets MUD\footnote{https://github.com/cablelabs/micronets-mud-tools}. NIST details the efficacy of these implementations against network-based attacks \cite{dodson2019securing}. Regarding the \mudfile{}s, \ensuremath{\mathsf{mudmaker}}\footnote{\href{https://www.mudmaker.org}{https://www.mudmaker.org}} is a web app specifically for creating \mudfile{}s. For devices that are not MUD-compliant, Hamza et al. created \ensuremath{\mathsf{MUDgee}} that uses the network traffic of the target IoT device to generate its \mudfile{} \cite{hamza2018clear}. Beside \mudvisualizer{}, which is the focus of this paper, \ensuremath{\mathsf{mudpp}}\footnote{https://github.com/iot-onboarding/mudpp} (MUD Pretty Printer) is another tool that is developed for summarizing the ACL in the \mudfile{}. However, since it does not perform any analysis on the interaction between the \mudfile{}s we did not consider it for our study. 


Usable access control has long been a challenge in usable security. An early study on the mitigation of human error in access control management was done by Maxion and Reeder \cite{maxion2005improving}. They found that visualization improves the rate of completing the assigned task by a factor of three. The error in these completed tasks was also reduced by up to 94\%. This study is particularly relevant to our work here because, like Maxion and Reeder, we selected computer and network science students with significant expertise.

The study conducted by Vaniea et al. \cite{vaniea2008access} also investigated the difficulty of translating policy rules into access control rules where they recommend visual feedback. They implemented SPARCLE \cite{vaniea2008access} to present the data in a table as a commonly used method of information visualization. The \ensuremath{\mathsf{Expandable Grid}} developed by Reeder et al. \cite{reeder2008expandable} for improving file permissions in Windows XP is another example in this category. 

Graph Visualization was previously used by \cite{KolomeetsAccessControlGraph} which is more similar to  \mudvisualizer's flow-based visualization \cite{andalibi2021mudviz}. 
Another study that concludes the importance of visualization is the work by Xu and colleagues \cite{xu2017system}. 
They investigate the uncertainties in access control decisions and found that a lack of feedback forced the administrators who intend to resolve access control conflicts into a trial and error mode. 
Moreover, Smetters et al. \cite{smetters2009users} found that limitations in the UI would lead to the reluctance to change the access control settings which applies to MUD deployment as well; manual evaluation of the interaction between multiple \mudfile{}s is a difficult and time-consuming task for system administrators. 

Erbenich et al. \cite{erbenich2019phishing} studied the efficacy of the link visualization to better protect the end-users against phishing. They break down the URL and only visualize the most critical part of it for successful phishing detection. 
The same concept was used in \mudvisualizer{} where only the summary of the \mudfile{}s was presented to the users. 
In another work, Scott and Ophoff \cite{scott2018investigating} conducted a user study to study the effectiveness of information security knowledge in decision making. 
By analyzing the knowledge-behavior gap, they found that a deeper technical understanding of cyber threats will help the user to effectively derive a more cautious and preventing behavior. 
This motivates one of our goals; to find out whether \mudvisualizer{} can help the users with higher knowledge and expertise in the analysis of the \mudfile{}s.

\section{Method}
\label{sec:method}
Our survey incorporated two groups of participants: the first group used \mudvisualizer{} for the analysis and the second group directly analyzed plain-text \mudfile{}s (hereinafter referred to as \ensuremath{\mathsf{mudviz}} and \ensuremath{\mathsf{plain}} groups respectively). 
The \texttt{plain} group acted as a control group to measure the efficacy of the \texttt{mudviz} group.
We asked a total of 81 questions, including three screening questions, five demographic questions, twenty-three questions related to the analysis of the \mudfile{}s (main experiment), forty expertise questions, and ten usability questions from the participants. 

Our \noindent\textbf{Screening Questionnaire and recruitment}
were designed to ensure that the participants have the required knowledge for analyzing a \mudfile{}. Before inclusion, participants had to show the knowledge of fundamentals of computer networking (i.e., understanding IP, Port, and access control) through manual parsing of components of a \mudfile. We focused on recruitment in an advanced computer networking course.

The \textbf{demographic questions} contained questions about age, gender, education, employment status, and income motivated from the study about the privacy for WEIRD populations \cite{henrich2010most}.

The core of the \textbf{experiment design} was 
 23 questions about the analysis of the \mudfile{}s. We first asked the participants about the remote servers or local devices allowed for a specific device given its \mudfile. This included two questions about the number of nodes devices allow-listed, seven questions about the name of these allowed nodes, and one question about between-node communication.  
 We also included thirteen questions about the Transport and Network layer protocols that are allow-listed for use, e.g. IP version, Port number, TCP vs UDP.


The \textbf{post-experiment} questions comprised 50 questions in two categories: forty expertise questions incorporating a set of computer expertise questions from \cite{rajivan2017factors} and ten usability questions from the System Usability Scale (SUS) \cite{brooke1996sus}.

\section{Results}
\label{sec:results}




31\% of our screening survey respondents (24 out of 76) failed to answer one or more of the screening questions and were not considered for the main study. The \textbf{participants} in our study were skewed with respect to gender (84.6\% male, 15.4\% female). Out of the total of 52 participants, 41 were below the age of 30 years. Over 70\% were students, with 50 participants having at least a technical Bachelors's degree. This includes only the participants who passed the screening questions. Participants were split equally between the two groups, \texttt{mudviz} and \texttt{plain}. 

In order to evaluate \textbf{participants' security and computer expertise}, they were presented with a set of 13 question categories. These questions were obtained from the set of computer expertise questions from \cite{rajivan2017factors}.
 For measures of \texttt{knowledge}, these were knowledge-based questions on (i) phishing (\texttt{Kphish}) (ii) certificates (\texttt{Kcert}) (iii) SQL commands (\texttt{Ksql}) (iv) intrusion detection systems (\texttt{Kids}) (v) port 80 (\texttt{K80}) (vi) Website markers for security (\texttt{Kweb}) (vii) defining IoT (\texttt{Kiot}) and (viii) access control (\texttt{Kac}). For single response questions, if the participants' answers matched the correct responses, these variables were coded as 1, otherwise 0. For multiple response questions (\texttt{Kphish} and \texttt{Kcert}), if the participants got a sum of correct values above the median in each category, the variables were coded as 1, otherwise 0. Since, all participants got responses to \texttt{Kiot} correct, these responses were removed in calculating the covariance matrix for factor analysis. 


We then performed a factor analysis on the remaining seven variables to create a \texttt{TotalKnowledge} variable. A scree plot and a test of hypothesis showed that a factor of one was sufficient to measure knowledge. This factor, \texttt{TotalKnowledge}, was a combination of four factors, calculated by the equation below:
\begin{center}
    $    TotalKnowledge 	\leftarrow (-0.5*Kcert) + (0.6*Ksql) + (0.6*Kids) + (0.7*K80)$
\end{center}


\noindent\texttt{TotalExperience} was similarly a combination of weighted factors, given by the equation below: 

\begin{center}
    $\indent TotalExperience \leftarrow  (0.5*Eyears) + (0.4*Elang) + (0.4*Efreq) $
\end{center}

That is, for the measure of experience, the remaining five questions on experience were evaluated - (i) prior computer expertise (\texttt{Eexp}) (ii) prior security expertise (\texttt{Etech}) (iii) programming languages known (\texttt{Elang}) (iv) years of experience working in security (\texttt{Eyears}) and (v) frequency of dealing with security problems (\texttt{Efreq}). Since the answers to these questions cannot be evaluated as correct/incorrect, we normalized each of the five variables and performed a second-factor analysis to create a \texttt{TotalExperience} variable. A scree plot and a test of hypothesis showed that a factor of one was sufficient to measure knowledge. 


We then evaluated the \textbf{Effect of Knowledge on Accuracy}
by first calculating \texttt{TotalKnowledge} and \texttt{TotalExperience}. \texttt{Accuracy} was measured as a summation of the correct answer to the 23 questions in the experiment, providing a raw accuracy percentage for each participant. 

\begin{figure}[!tbp]
  \centering
  \subfloat[Regression for \texttt{mudviz} group]{\includegraphics[width=0.5\textwidth]{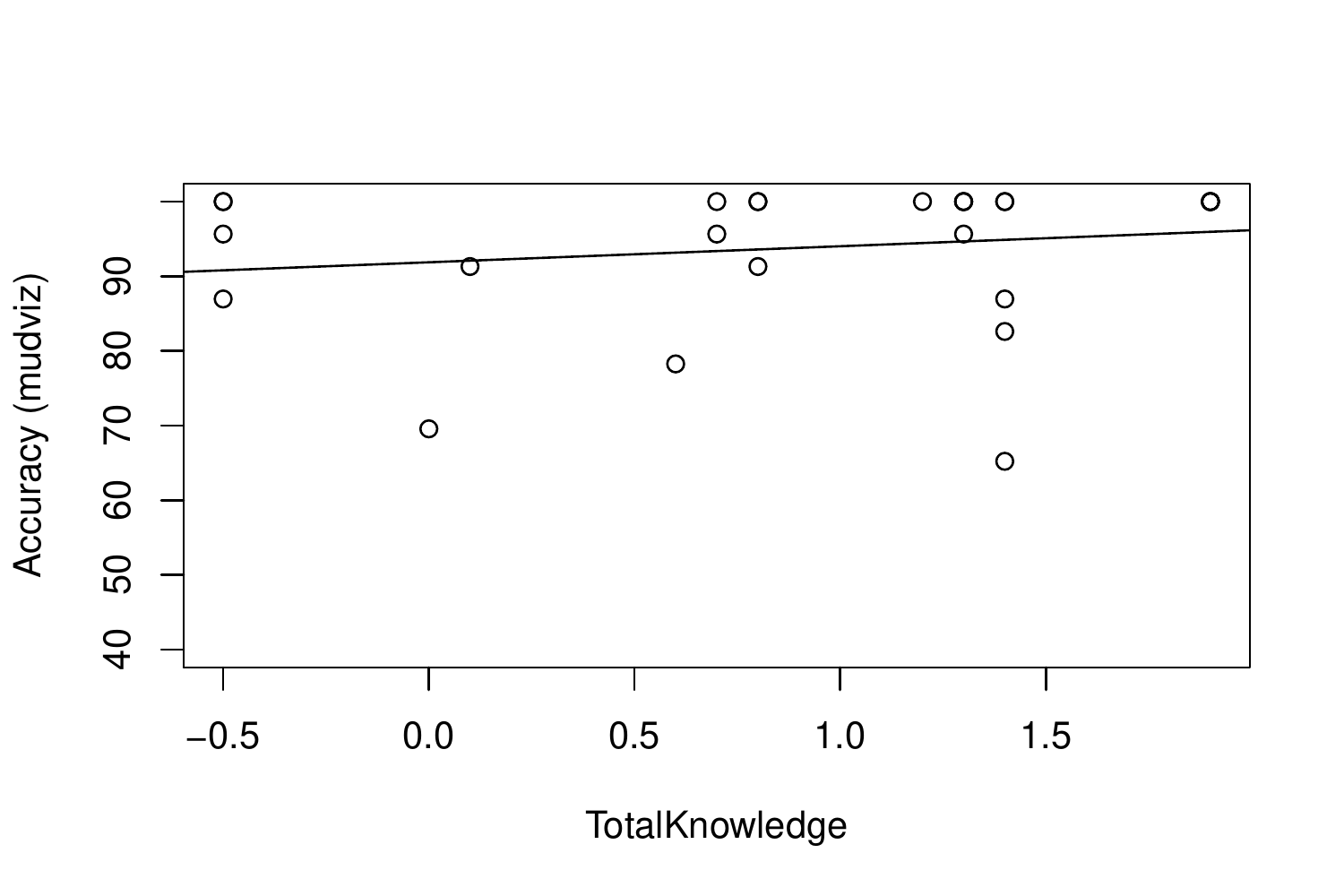}\label{fig:1a}}
  \hfill
  \subfloat[Regression for \texttt{plain} group]{\includegraphics[width=0.5\textwidth]{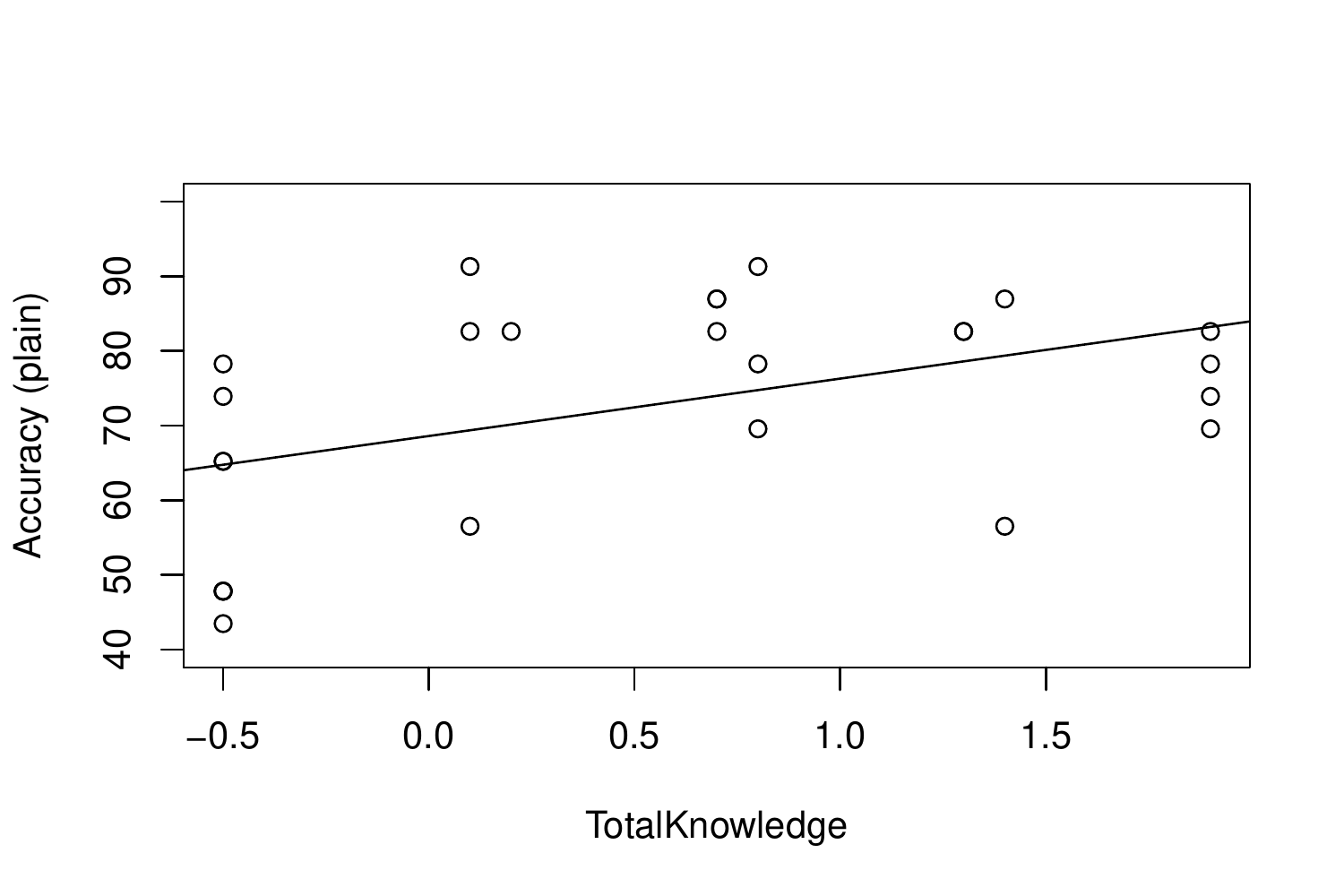}\label{fig:1b}}
  \qquad%
  \subfloat[\texttt{Accuracy} v.\texttt{TotalKnowledge}(\texttt{mudviz})]{\label{fig:c}\includegraphics[width=0.5\textwidth]{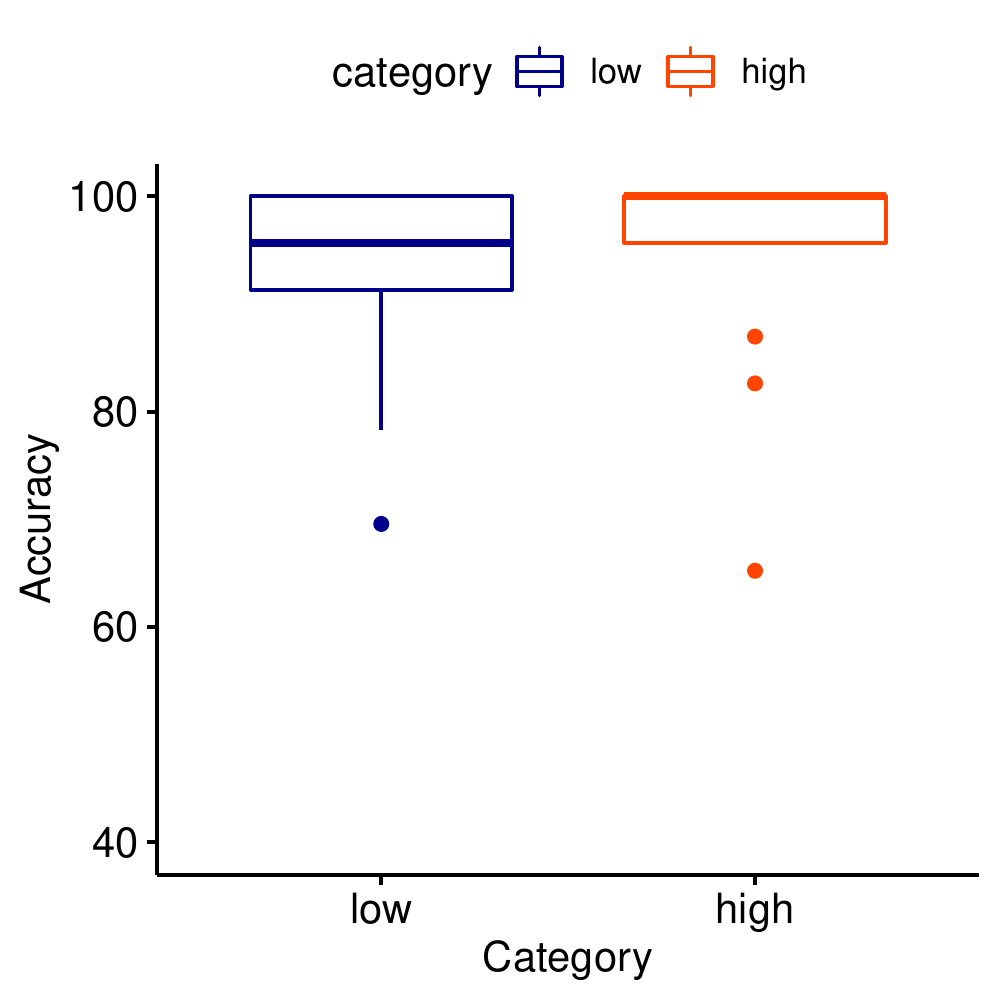}}
\subfloat[\texttt{Accuracy} v.\texttt{TotalKnowledge}(\texttt{plain})]{\label{fig:d}\includegraphics[width=0.5\textwidth]{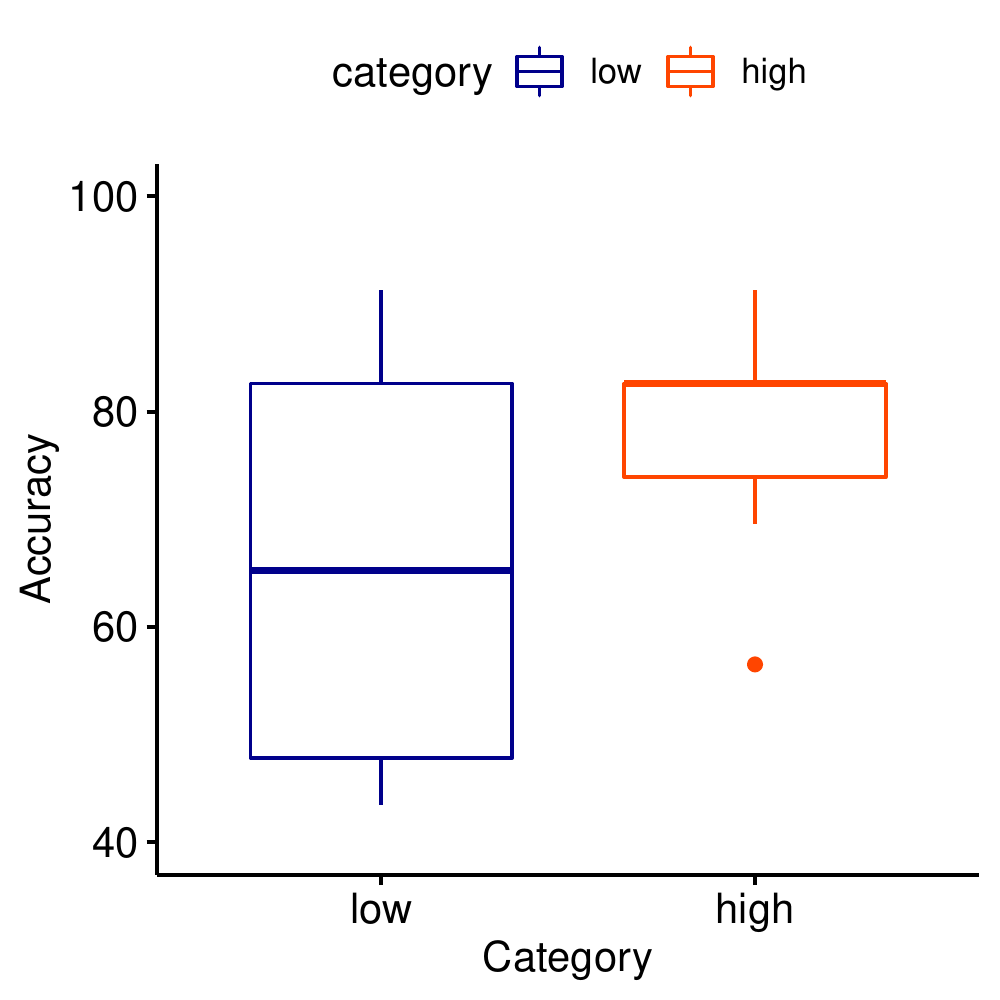}}%
\caption{ 
(a) and (b) show the scatter plot of \texttt{Accuracy} against \texttt{TotalKnowledge}. 
(c) and (d) show \texttt{Accuracy} for four groups, indicating that the effect of the \mudvisualizer{} is consistently positive across knowledge groups.}
\label{fig:regknow}
\end{figure}

In order to answer \textbf{RQ1}, we first performed a linear regression to measure the effect of the independent variable \texttt{TotalKnowledge} on the dependent variable \texttt{Accuracy} for both groups (Fig. \ref{fig:1a} and \ref{fig:1b}). 
Unsurprisingly, knowledge has a positive effect on the accuracy of the analysis of the \mudfiles{}. We also found that the effect of \texttt{TotalKnowledge} on \texttt{Accuracy} is significant in the \texttt{plain} group ($b = 7.689, p-value = 0.0164$) but not for the \texttt{mudviz} group ($b = 2.148, p-value = 0.406$). 
Thus, participants in the \texttt{mudviz} group seemed to have the same level of accuracy across computer and security knowledge levels. 
However, this is not the case for plain text files. Participants with greater \texttt{TotalKnowledge} seemed to have significantly high \texttt{Accuracy} in the \texttt{plain} group. This suggests that normally a high level of security expertise is needed to understand textual \mudfiles{}, but that an effective visualization can result in accuracy by moderate experts indistinguishable from that of the most expert.

The results of a linear regression conducted on each of the factors indicate that none of the factors in the \texttt{mudviz} have a significant effect on \texttt{Accuracy}, but some factors in the \texttt{plain} group are significant. Table \ref{tab:regknowind} shows the regression of individual knowledge factors for both groups. We see that the \texttt{Kphish}, \texttt{Kids}, \texttt{K80}, and, \texttt{Ksql} are more strongly significant than the other factors in contributing to \texttt{Accuracy}. 
\begin{table}[b]
\caption{Regression analysis for individual knowledge factors versus accuracy in MUD analysis (showing significant components only).}
\centering
\begin{tabular}{@{}lllll@{}}
\toprule
\multirow{2}{*}{Factors} & \multicolumn{2}{c}{mudviz} & \multicolumn{2}{c}{plain}           \\ \cmidrule(l){2-5} 
                         & co-efficient   & p-value   & co-efficient    & p-value           \\ \cmidrule(r){1-1}
Kphish                   & 7.412          & 0.136     & \textbf{12.847} & \textbf{0.0445 *} \\
Kids                     & 1.967          & 0.624     & \textbf{11.594} & \textbf{0.0413 *} \\
K80                      & 3.370          & 0.411     & \textbf{9.576}  & \textbf{0.0968 .} \\
Ksql                     & 1.733          & 0.701     & \textbf{11.957} & \textbf{0.0348 *} \\
\bottomrule
TotalKnowledge           & 2.148          & 0.406     & \textbf{7.689}  & \textbf{0.0164 *} \\ \bottomrule
\end{tabular}

\label{tab:regknowind}
\end{table}

To answer the first part of \textbf{RQ3}, we analyzed whether \texttt{TotalKnowledge} can be divided into sub-groups of knowledge and expertise respectively; and how these interact with \texttt{Accuracy}. We sorted the participants from each of the \texttt{mudviz} and \texttt{plain} groups in ascending order based on their \texttt{TotalKnowledge} 
with 13 participants in each sub-group. A signed Wilcoxon Rank-sum test indicated significant difference between 
the four sub-group categories, with p-values between the low and high groups of less than $0.001$. We conducted an ordinal logistic regression between the two categories (low and high) for each of the two groups for \texttt{TotalKnowledge} against \texttt{Accuracy}, (a) \texttt{Mudviz} and (b) \texttt{Plain}. As seen in Fig. \ref{fig:c}, the accuracy in correct interpretation of the \mudfile{}s did not vary significantly between high and low knowledge categories in the \texttt{Mudviz} group ($b = -0.018, p-value = 0.663$). However, in case of the \texttt{plain group} (Fig. \ref{fig:d}) \texttt{TotalKnowledge} played a significant role in increasing the accuracy ($b = -0.066, p-value = 0.054$).  The accuracy was consistently higher in the \texttt{mudviz} group compared to the \texttt{plain} group in all cases. 

\begin{figure}[t]
  \centering
  \subfloat[Regression for \texttt{mudviz} group]{\includegraphics[width=0.45\textwidth]{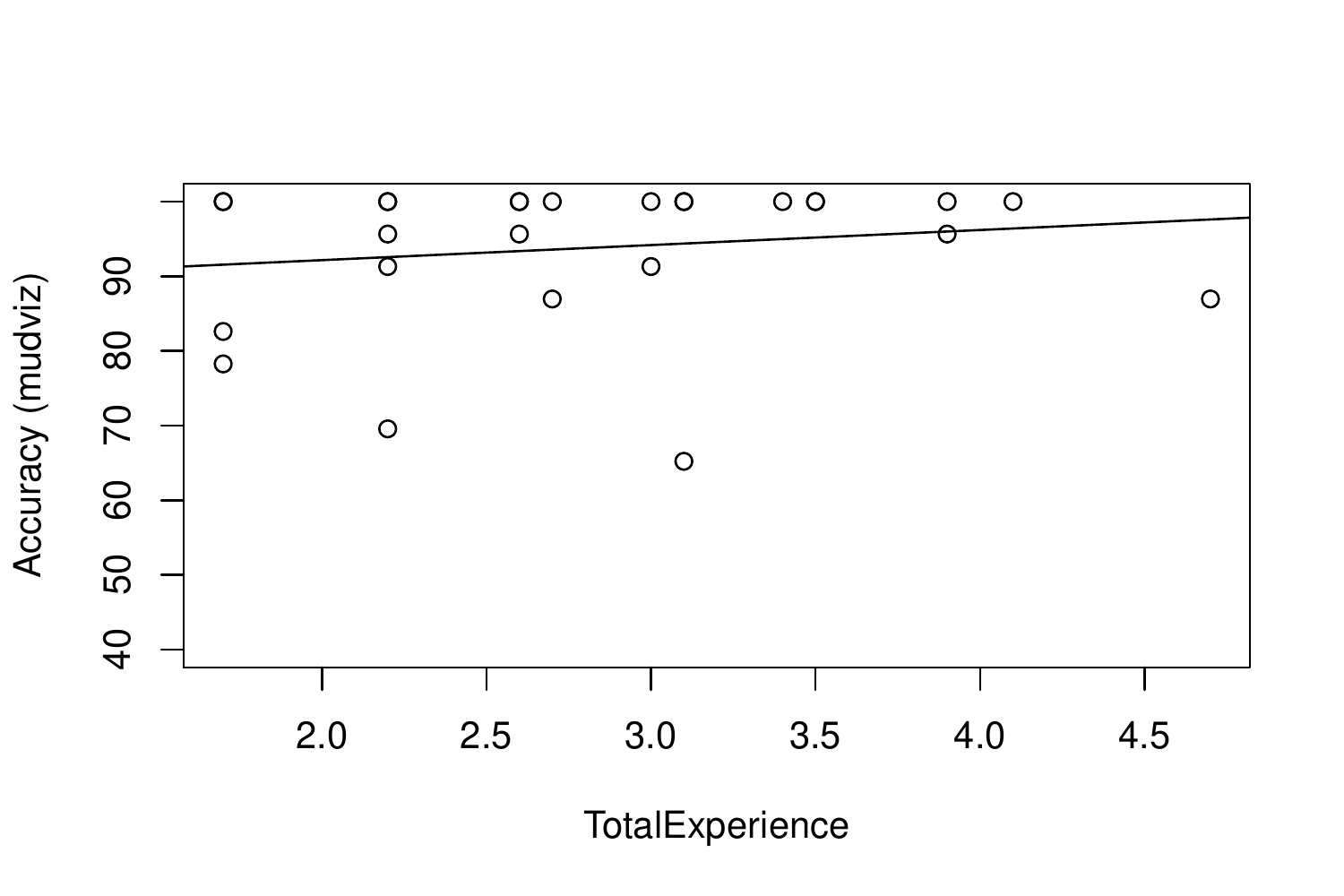}\label{fig:2a}}
  \hfill
  \subfloat[Regression for \texttt{plain} group]{\includegraphics[width=0.45\textwidth]{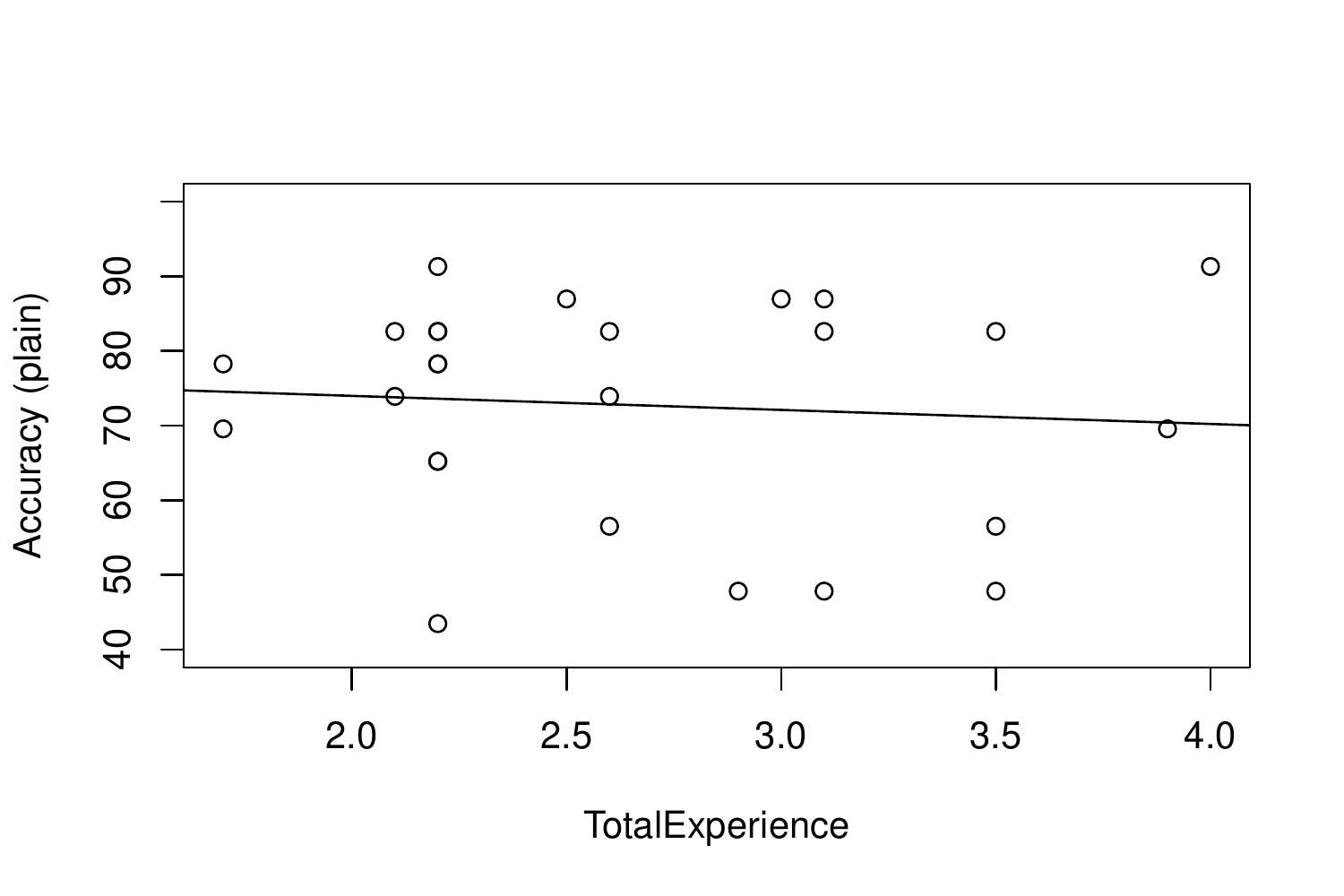}\label{fig:2b}}\qquad
  \subfloat[\texttt{Accuracy} v.\texttt{TotalExperience} (\texttt{mudviz})]{\label{fig:a}\includegraphics[width=0.45\linewidth]{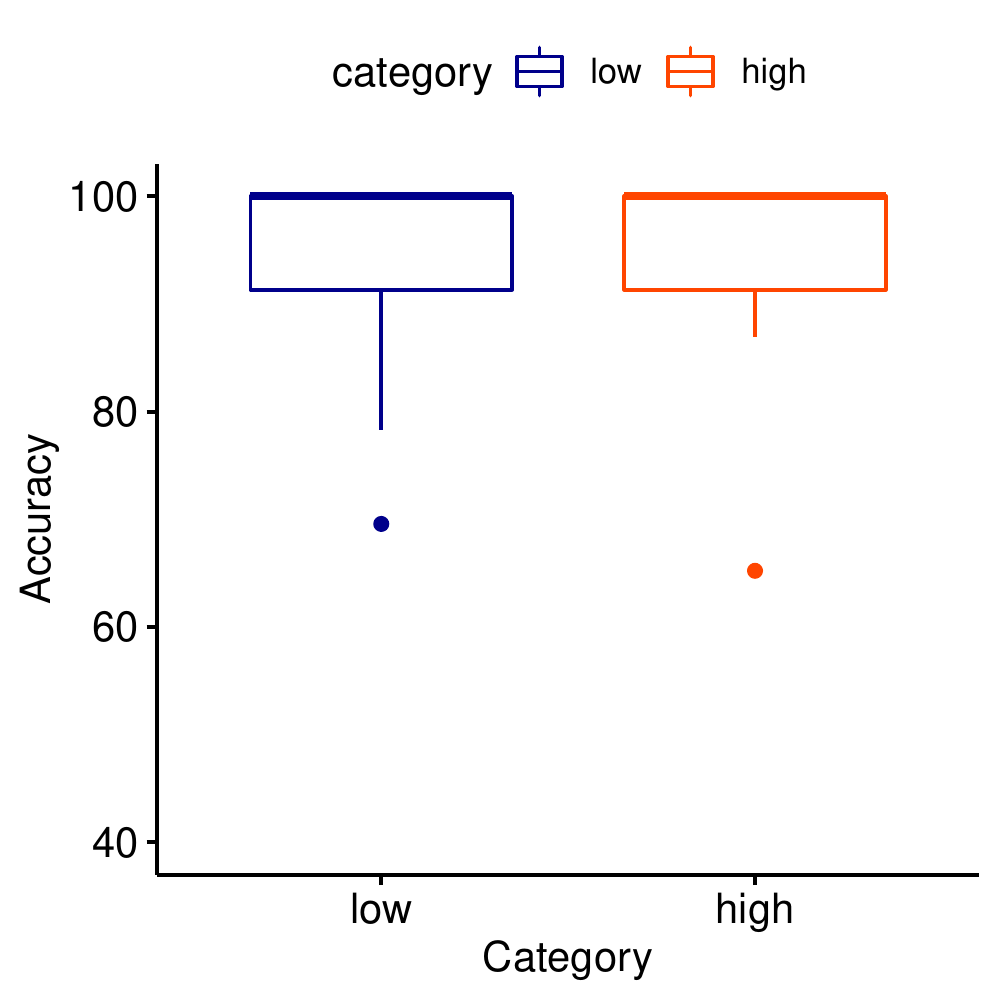}}
\subfloat[\texttt{Accuracy} v.\texttt{TotalExperience} (\texttt{plain})]{\label{fig:b}\includegraphics[width=0.45\linewidth]{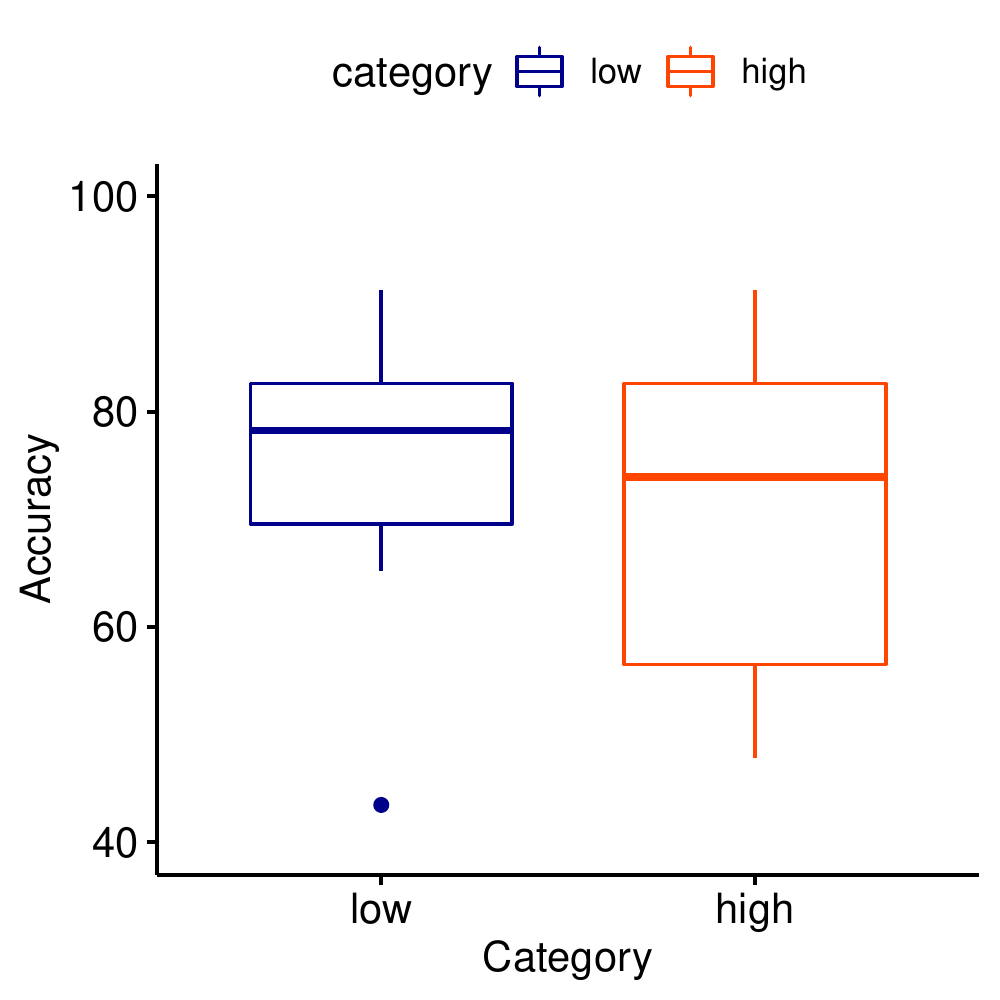}}\\
  \caption{ 
  (a) and (b) show the scatter plot of \texttt{Accuracy} against \texttt{TotalExperience}. 
  (c) and (d) show \texttt{Accuracy} for four groups, indicating that the effect of the \mudvisualizer{} is consistently positive across all experience groups.}
\label{fig:regexp}
\end{figure}

\begin{table}[b]
\caption{Regression analysis for individual experience factors versus accuracy in MUD analysis.}
\centering
\begin{tabular}{@{}lllll@{}}
\toprule
\multirow{2}{*}{Factors} & \multicolumn{2}{c}{mudviz} & \multicolumn{2}{c}{plain}            \\ \cmidrule(l){2-5} 
                         & co-efficient   & p-value   & co-efficient   & p-value             \\ \cmidrule(r){1-1}
Eexp                     & 1.187          & 0.393     & \textbf{6.505} & \textbf{0.00299 **} \\
Efreq                   & 1.789          & 0.259     & -4.797         & 0.18                \\
Eyears                   & 0.345          & 0.899     & -0.050         & 0.989               \\
TotalExperience          & 2.018          & 0.425     & -1.879         & 0.687               \\ \bottomrule
\end{tabular}

\label{tab:regexpind}
\end{table}

To investigate \textbf{Effect of Experience on Accuracy}
(\textbf{RQ2}) we began with a linear regression to measure the effect of independent variable \texttt{TotalExperience} on \texttt{Accuracy} for the both groups. Fig. \ref{fig:2a} and \ref{fig:2b} show the scatterplot and the regression lines for each of the \texttt{mudviz} and \texttt{plain} groups respectively. Unsurprisingly, experience has a positive effect on \texttt{Accuracy} in case of \texttt{mudviz}.
Yet there appears to a weak negative effect on \texttt{Accuracy} in case of \texttt{plain} in Fig. \ref{fig:b}, which we delve into in Table~\ref{tab:regexpind} below.

\texttt{TotalExperience} is not significant for \texttt{Accuracy} in either case of the \texttt{plain} group ($b = -1.879, p-value = 0.687$) or the \texttt{mudviz} group ($b = 2.018, p-value = 0.425$); although differences in the distribution of the \texttt{plain} are apparent. Thus, participants in the group that were presented with the \mudvisualizer{} seemed to have the same level of accuracy across computer and security experience levels.

Taking a closer look at the experience factors by conducting a linear regression for each of the factors, we see that none of the factors in the \texttt{mudviz} group affect \texttt{Accuracy} significantly, but the \texttt{Eexp} factor in the \texttt{plain} group does. In that case, \texttt{Eexp} is significant and positive.
Table \ref{tab:regexpind} shows the regression of individual experience factors for both groups.
\texttt{Eexp} is a set of Booleans from querying if participants had experience with any of the following: designing a website, registering a domain name, using SSH, configuring a firewall, creating a database, installing a computer program, and writing a computing program. 
The intriguing but not significant negative effect on \texttt{Accuracy} is due to \texttt{Efreq} (frequency of handling security incidents) and \texttt{Eyears} (years of experience working in the security field). It is possible that this may result from less experienced people defining security incidents (e.g., spam vs. an intrusion) or being in the security field differently (e.g., total years in coursework vs. years in incident response not DevOps).

To answer the second part of the \textbf{RQ3}, we analyzed whether \texttt{TotalExperience} can be divided into sub-groups of knowledge and expertise respectively, and how they affect the \texttt{Accuracy}. We sorted the participants from each of the \texttt{mudviz} and \texttt{plain} groups in ascending order based on their \texttt{TotalExperience}. 
Again, we considered 13 participants in each sub-group. A signed Wilcoxon Rank-sum test showed that 
the four sub-group categories are significantly different, with $p-values$ between each of the low versus high groups being less than $0.001$. We conducted an ordinal logistic regression between the two categories (low and high) for each of the two groups of \texttt{TotalExperience} against \texttt{Accuracy}, (a) \texttt{Mudviz} (b) \texttt{Plain}. The results illustrated that for \texttt{Mudviz} ($b = 2.018, p-value = 0.425$), the accuracy in interpreting the \mudfile{} correctly was the nearly the same for low and high \texttt{TotalExperience} (Similar to \texttt{TotalKnowledge}).
\section{Conclusions}
\label{sec:conclusion}
In this work, we sought to evaluate the efficacy of \mudvisualizer{} for correct evaluation of \mudfile{} by participants with some expertise.  We report on the increase in efficacy among all participants, showing that the difference in the performance of network engineers with and without knowledge of security or security expertise was significant. More-so, accuracy of participants using the \mudvisualizer{} showed knowledge of security to be insignificant (among these participants).  Given the difficulty of providing network engineers with security expertise, having a visualization that decreases the cost of inexperience argues for the importance of human factors in standards. Beyond that we found evidence that interpretation of security questions may be having a subtle impact on the results; those with less experience may not be reporting experience with the same baseline as those with more. This phenomena is worthy of additional research, although in this case any impact would strengthen the results.


\begin{acknowledgements}
This research was supported in part by the National Science Foundation awards CNS 1565375 and CNS 1814518, as well as the grant \#H8230-19-1-0310, Cisco Research Support, Google Research, and the Comcast Innovation Fund. Any opinions, findings, and conclusions, or recommendations expressed in this material are those of the author(s) and do not necessarily reflect the views of the National Science Foundation, Cisco, Comcast, Google, nor Indiana University.

\end{acknowledgements}

\bibliographystyle{spmpsci}      

\bibliography{refs}

\end{document}